\documentclass[english]{article}
\usepackage[T1]{fontenc}
\usepackage[latin9]{inputenc}
\usepackage{amsmath}
\usepackage{amssymb}
\usepackage{graphicx}
\usepackage{esint}

\makeatletter
\newcommand{\lyxaddress}[1]{
\par {\raggedright #1
\vspace{1.4em}
\noindent\par}
}

\makeatother

\usepackage{babel}
\begin{document}

\title{\textbf{\large Average quantum dynamics of closed systems over stochastic
Hamiltonians}}

\author{Li Yu and Daniel F. V. James}

\maketitle

\lyxaddress{}

\lyxaddress{\begin{center}
\emph{Department of Physics, University of Toronto, 60 St. George
Street, Toronto, Ontario M5S 1A7, Canada} 
\par\end{center}}
\begin{abstract}
{\normalsize We develop a master equation formalism to describe the
evolution of the average density matrix of a closed quantum system
driven by a stochastic Hamiltonian. The average over random processes
generally results in decoherence effects in closed system dynamics,
in addition to the usual unitary evolution. We then show that, for
an important class of problems in which the Hamiltonian is proportional
to a Gaussian random process, the 2nd-order master equation yields
exact dynamics. The general formalism is applied to study the examples
of a two-level system, two atoms in a stochastic magnetic field and
the heating of a trapped ion.}{\normalsize \par}

\end{abstract}
\begin{flushright}
PACS number(s): 03.65.Ca, 03.65.Yz, 05.40.-a, 03.67.Lx 
\par\end{flushright}

\section{Introduction}

The density operator encapsulates all the statistical information
about the state of a quantum system. The evolution of the density
operator of a closed system is governed by the Hamiltonian. In practice,
the Hamiltonian can seldom be strictly determined or precisely controlled
-- it fluctuates both in a temporal sense and between repeated realizations,
which can be mathematically described by random processes. Therefore,
instead of treating any Hamiltonian as deterministic in an idealized
manner, we would like to take such fluctuations into account explicitly
when studying quantum dynamics. Our goal is to obtain the \emph{average}
dynamics in the following sense: Suppose an ensemble of systems are
prepared in some initial state and subsequently evolve under a randomly
fluctuating Hamiltonian, how does the density matrix that describes
the ensemble as a whole evolve?

Previous work on stochastic average dynamics was done by Budini \cite{budini}
using a variational calculus method and Novikov's theorem, and by
Guha et al. \cite{guha} using a non-perturbative cluster cumulant
method. A different kind of average dynamics over the time domain
was studied by Gamel and James \cite{gamel}, assuming the deterministic
(i.e. non-stochastic) Hamiltonian but taking into account the finite
time-window of measurements. The master equation formalism is also
widely used in the study of open systems dynamics \cite{breuer}.
It should be noted that, despite the formal similarity, our study
is on the dynamics of \emph{closed} quantum systems and no environment
is involved.

In this paper, we will first adopt a series expansion approach and
derive a time-local master equation that describes the ensemble-average
dynamics of a general quantum system. The general formalism is then
used to study a representative class of Hamiltonians obeying Gaussian
statistics. Finally, we apply the master equation method to some physical
examples and find interesting phenomena such as fluctuation-induced
decoherence and decoherence-induced disentanglement. Throughout, our
results are compared to exact dynamics and the validity of the master
equation approach is discussed.

\section{Theory}

\subsection{Ensemble-average density matrix}

Consider a closed, but not isolated, system for which the Hamiltonian
is determined by some classical stochastic quantity $x(t)$. Suppose
an experiment is carried out repeatedly with each realization labelled
by $\mu$. The evolution of the density matrix $\rho^{\mu}(t)$ that
describes the quantum system in the $\mu$-th realization is governed
by the Hamiltonian $\hat{H}^{\mu}(t)=\hat{H}[x^{\mu}(t)]$, and is
given by

\begin{equation}
\rho^{\mu}(t)=\hat{U}^{\mu}(t,t_{0})\rho_{0}\hat{U}^{\mu\dagger}(t,t_{0}),
\end{equation}
 where $\rho_{0}$ is the initial density matrix, which is assumed
to be uncorrelated with $x(t)$ and thus is the same in all realizations.
The unitary evolution operator $\hat{U}^{\mu}(t,t_{0})$ obeys the
equation of motion,

\begin{equation}
i\hbar\frac{\partial}{\partial t}\hat{U}^{\mu}(t,t_{0})=\hat{H}^{\mu}(t)\hat{U}^{\mu}(t,t_{0}).\label{eq:eom}
\end{equation}

The average density matrix $\overline{\rho}(t)$ is defined as follows,

\begin{equation}
\overline{\rho}(t)\equiv\underset{N\rightarrow\infty}{\lim}\frac{1}{N}\overset{N}{\underset{\mu=1}{\sum}}\rho^{\mu}(t).
\end{equation}
 It can be shown that $\overline{\rho}(t)$ is Hermitian, positive
and of unit trace, which is ensured by the properties of the individual
density matrices $\rho^{\mu}(t)$. Thus the operator $\overline{\rho}(t)$
is indeed a physical density matrix, describing the average statistics
of the ensemble of realizations as a whole.

The equation of motion for $\overline{\rho}(t)$ is formally given
by

\begin{equation}
i\hbar\frac{\partial}{\partial t}\overline{\rho}(t)=\underset{N\rightarrow\infty}{\lim}\frac{1}{N}\overset{N}{\underset{\mu}{\sum}}i\hbar\frac{\partial}{\partial t}\rho^{\mu}(t)=\underset{N\rightarrow\infty}{\lim}\frac{1}{N}\overset{N}{\underset{\mu}{\sum}}[\hat{H}^{\mu}(t),\rho^{\mu}(t)]=\overline{[\hat{H}(t),\rho(t)]}.
\end{equation}
 However, since the right hand side cannot be written as a function
of $\overline{\rho}(t)$, the equation is not of a closed form and
thus not very useful. With the goal of obtaining a closed equation
for $\overline{\rho}(t)$, we resort to a series expansion approach.

\subsection{Series expansion of the evolution operator}

Following the standard recipe for perturbative expansion \cite{shankar},
the unitary operator $\hat{U}^{\mu}(t,t_{0})$ in a particular realization
$\mu$ can be written as a power series in $\lambda$ (a parameter
controlling the {}``strength'' of the Hamiltonian): 
\begin{equation}
\hat{U}^{\mu}(t,t_{0})=\overset{\infty}{\underset{n=0}{\sum}}\lambda^{n}\hat{U}_{n}^{\mu}(t,t_{0})
\end{equation}
 where 
\begin{eqnarray}
\hat{U}_{0}^{\mu}(t,t_{0}) & = & \hat{I},\\
\hat{U}_{n}^{\mu}(t,t_{0}) & = & \frac{1}{i\hbar}\intop_{t_{0}}^{t}dt'\hat{H}^{\mu}(t')\hat{U}_{n-1}^{\mu}(t',t_{0}),\, n\geqslant1.
\end{eqnarray}

Thus $\overline{\rho}(t)$ can be expressed in terms of $\hat{U}_{n}^{\mu}(t,t_{0})$
and $\lambda$:

\begin{multline}
\overline{\rho}(t)=\overline{\left(\underset{m}{\sum}\lambda^{m}\hat{U}_{m}(t,t_{0})\right)\rho_{0}\left(\underset{n}{\sum}\lambda^{n}\hat{U}_{n}^{\dagger}(t,t_{0})\right)}=\overset{\infty}{\underset{k=0}{\sum}}\lambda^{k}\mathcal{E}_{k}[\rho_{0}]\equiv\mathcal{E}[\rho_{0}],\label{eq:map}
\end{multline}
 where $\mathcal{E}_{k}[\rho_{0}]$ is the time-dependent map defined
as 
\begin{equation}
\mathcal{E}_{k}[\rho_{0}]\equiv\overset{k}{\underset{j=0}{\sum}}\overline{\hat{U}_{k-j}(t,t_{0})\rho_{0}\hat{U}_{j}^{\dagger}(t,t_{0})},\label{eq:define maps}
\end{equation}
 and $\mathcal{E}[\rho_{0}]$ is a completely positive linear map
\cite{choi}. Although their argument is a density matrix in this
instance, $\mathcal{E}_{k}$ and $\mathcal{E}$ can act on any operator
in general.

\subsection{Inverse transformation}

The map $\mathcal{E}$ is a linear transformation that maps $\rho_{0}$
to $\overline{\rho}$. Since both $\rho_{0}$ and $\overline{\rho}$
are operators in the same Hilbert space, and thus of the same dimension,
it is natural to postulate that an inverse transformation $\mathcal{F}=\mathcal{E}^{-1}$
exists that maps $\overline{\rho}$ to $\rho_{0}$. That is,

\begin{equation}
\rho_{0}=\mathcal{E}^{-1}[\overline{\rho}]\equiv\mathcal{F}[\overline{\rho}].\label{eq:inverse map}
\end{equation}
 Note that the meaning of {}``inverse'' is purely mathematical here:
the map $\mathcal{E}^{-1}$ is not to be confused with an inverse
dynamical evolution in the physical sense. According to the semigroup
property, a completely positive, trace preserving (CPTP) linear map
is physically invertible if and only if it is an unitary map (see
Section 3.8 of Ref.\cite{preskill}). Therefore, in general, a CPTP
map $\mathcal{E}:\,\rho_{0}\rightarrow\overline{\rho}$ does not have
a physical inverse, that is, we cannot find another CPTP map that
gives $\overline{\rho}\rightarrow\rho_{0}$. However, the mathematical
inverse $\mathcal{E}^{-1}$ that serves our purpose here needs not
be CPTP.

Since the composition of a transformation and its inverse is the identity
transformation, the identity $\mathcal{F}[\mathcal{E}[\rho]]=\mathcal{I}[\rho]$
holds for an arbitrary operator $\rho$. Following \cite{gamel},
we adopt the ansatz that $\mathcal{F}$ can be expanded in powers
of $\lambda$, $\mathcal{F}[\rho]=\overset{\infty}{\underset{m=0}{\sum}}\lambda^{m}\mathcal{F}_{m}[\rho]$.
Then we have

\begin{equation}
\overset{\infty}{\underset{m=0}{\sum}}\lambda^{m}\mathcal{F}_{m}[\overset{\infty}{\underset{n=0}{\sum}}\lambda^{n}\mathcal{E}_{n}[\rho]]=\overset{\infty}{\underset{k=0}{\sum}}\lambda^{k}\left(\overset{k}{\underset{j=0}{\sum}}\mathcal{F}_{j}[\mathcal{E}_{k-j}[\rho]]\right)=\lambda^{0}\mathcal{I}[\rho].
\end{equation}
 Collecting terms of like powers in $\lambda$, we obtain the set
of equations involving $\mathcal{F}_{m}$ and $\mathcal{E}_{n}$:

\begin{eqnarray}
\mathcal{F}_{0}[\mathcal{E}_{0}[\rho]] & = & \mathcal{I}[\rho],\\
\mathcal{F}_{0}[\mathcal{E}_{1}[\rho]] & + & \mathcal{F}_{1}[\mathcal{E}_{0}[\rho]]=0,\\
\mathcal{F}_{0}[\mathcal{E}_{2}[\rho]] & + & \mathcal{F}_{1}[\mathcal{E}_{1}[\rho]]+\mathcal{F}_{2}[\mathcal{E}_{0}[\rho]]=0,
\end{eqnarray}
 and so on. Solving for $\mathcal{F}_{m}$ in terms of $\mathcal{E}_{n}$,
and making use of $\mathcal{E}_{0}=\mathcal{I}$ as defined in Eq.(\ref{eq:define maps}),
we have

\begin{eqnarray}
\mathcal{F}_{0}[\rho] & = & \mathcal{E}_{0}[\rho]=\mathcal{I}[\rho],\\
\mathcal{F}_{1}[\rho] & = & -\mathcal{E}_{1}[\rho],\\
\mathcal{F}_{2}[\rho] & = & -\mathcal{E}_{2}[\rho]+\mathcal{E}_{1}[\mathcal{E}_{1}[\rho]],
\end{eqnarray}
 and so on.

\subsection{Master equation}

Differentiating Eq.(\ref{eq:map}) with respect to time and making
use of the inverse relation in Eq.(\ref{eq:inverse map}), we obtain
the following equation: 
\begin{equation}
i\hbar\frac{\partial}{\partial t}\overline{\rho}(t)=i\hbar\dot{\mathcal{E}}[\rho_{0}]=i\hbar\dot{\mathcal{E}}[\mathcal{F}[\overline{\rho}(t)]].
\end{equation}
 Here, the notation $\dot{\mathcal{E}}[\rho]$ means first taking
time-derivative of the time-dependent transformation $\mathcal{E}$
to obtain a new transformation denoted by $\dot{\mathcal{E}}$ and
then letting $\dot{\mathcal{E}}$ act on $\rho$; the argument $\rho$
is not differentiated whether or not it is time-dependent. Assuming
the order of differentiation and summation can be switched, we have

\begin{equation}
\dot{\mathcal{E}}[\mathcal{F}[\overline{\rho}(t)]]=\overset{\infty}{\underset{n=0}{\sum}}\lambda^{n}\dot{\mathcal{E}_{n}}[\overset{\infty}{\underset{m=0}{\sum}}\lambda^{m}\mathcal{F}_{m}[\overline{\rho}(t)]]=\overset{\infty}{\underset{k=0}{\sum}}\lambda^{k}\left(\overset{k}{\underset{j=0}{\sum}}\dot{\mathcal{E}_{j}}[\mathcal{F}_{k-j}[\overline{\rho}(t)]]\right);
\end{equation}
 thus the equation of motion can be written as

\begin{equation}
i\hbar\frac{\partial}{\partial t}\overline{\rho}(t)=\overset{\infty}{\underset{k=0}{\sum}}\lambda^{k}\left(i\hbar\overset{k}{\underset{j=0}{\sum}}\dot{\mathcal{E}_{j}}[\mathcal{F}_{k-j}[\overline{\rho}(t)]]\right)\equiv\overset{\infty}{\underset{k=0}{\sum}}\lambda^{k}\mathcal{L}_{k}[\overline{\rho}(t)].\label{eq:me-series expansion}
\end{equation}
 Evaluating $\mathcal{F}_{m}$ and $\dot{\mathcal{E}_{n}}$ explicitly,
we find

\begin{eqnarray}
\mathcal{L}_{0}[\rho] & = & i\hbar\dot{\mathcal{E}_{0}}[\mathcal{F}_{0}[\rho]]=0,\\
\mathcal{L}_{1}[\rho] & = & i\hbar\dot{\mathcal{E}_{0}}[\mathcal{F}_{1}[\rho]]+i\hbar\dot{\mathcal{E}_{1}}[\mathcal{F}_{0}[\rho]]=\overline{\hat{H}}\rho-\rho\overline{\hat{H}},\\
\mathcal{L}_{2}[\rho] & = & i\hbar\dot{\mathcal{E}_{0}}[\mathcal{F}_{2}[\rho]]+i\hbar\dot{\mathcal{E}_{1}}[\mathcal{F}_{1}[\rho]]+i\hbar\dot{\mathcal{E}_{2}}[\mathcal{F}_{0}[\rho]]\nonumber \\
 & = & \overline{\hat{H}\hat{U_{1}}}\rho-\overline{\hat{H}}\,\overline{\hat{U}_{1}}\rho+\overline{\hat{H}\rho\hat{U}_{1}^{\dagger}}-\overline{\hat{H}}\rho\overline{\hat{U}_{1}^{\dagger}}\nonumber \\
 &  & -\rho\overline{\hat{U}_{1}^{\dagger}\hat{H}}+\rho\overline{\hat{U}_{1}^{\dagger}}\,\overline{\hat{H}}-\overline{\hat{U}_{1}\rho\hat{H}}+\overline{\hat{U}_{1}}\rho\overline{\hat{H}},\label{eq:L(2)}
\end{eqnarray}
 and so on. Note again that the argument $\rho$ is not to be averaged
or differentiated and that terms like $\overline{\hat{H}}$ are time-dependent
just as $\mathcal{L}_{k}[\rho]$ are time-dependent transformations.

Keeping terms up to 2nd order and setting $\lambda=1$ in Eq.(\ref{eq:me-series expansion}),
a time-local master equation is thus obtained for the evolution of
$\overline{\rho}(t)$:

\begin{equation}
i\hbar\frac{\partial}{\partial t}\overline{\rho}(t)=[\overline{\hat{H}},\overline{\rho}(t)]+\hat{A}\overline{\rho}(t)-\overline{\rho}(t)\hat{A}^{\dagger}+\mathcal{D}[\overline{\rho}(t)],
\end{equation}
 where $\hat{A}\equiv\overline{\hat{H}\hat{U}_{1}}-\overline{\hat{H}}\,\overline{\hat{U}_{1}}$
and $\mathcal{D}[\rho]\equiv\overline{\hat{H}\rho\hat{U}_{1}^{\dagger}}-\overline{\hat{H}}\rho\overline{\hat{U}_{1}^{\dagger}}-\overline{\hat{U}_{1}\rho\hat{H}}+\overline{\hat{U}_{1}}\rho\overline{\hat{H}}$.
The effective Hamiltonian responsible for unitary evolution is

\begin{equation}
\hat{H}_{eff}\equiv\overline{\hat{H}}+\frac{1}{2}(\hat{A}+\hat{A}^{\dagger}),\label{eq:effective H}
\end{equation}
 with which the master equation can be written in a more insightful
way,

\begin{equation}
i\hbar\frac{\partial}{\partial t}\overline{\rho}(t)=[\hat{H}_{eff},\overline{\rho}(t)]+\frac{1}{2}\left\{ \hat{A}-\hat{A}^{\dagger},\overline{\rho}(t)\right\} +\mathcal{D}[\overline{\rho}(t)].\label{eq:general me}
\end{equation}
 It can be shown that the right-hand side of the equation can be put
into the Lindblad form, which ensures Hermiticity, complete positivity
and trace preservation of the evolution.

This result is formally similar to a previous work on average dynamics
\cite{gamel}. However, the physical meaning is different since the
derivation in that case is for a time-average density matrix in a
\emph{single} realization. Incidentally, our result may also be reminiscent
of some master equations for the reduced density matrix of open quantum
systems. But it should be emphasized that our derivation is for a
closed system and thus quantum entanglement with environment does
not play a role here.

Note that the above results are formally applicable to an interaction-picture
density matrix, though we implicitly assume the Schrödinger picture
in the derivation. The only difference is in the interpretation of
the density matrix: When we use the interaction-picture density matrix
$\overline{\rho}_{I}$, the expectation value of an observable $\hat{O}$
is given by $\langle\hat{O}\rangle=Tr\left(\hat{O}_{I}\overline{\rho}_{I}\right)$,
where $\hat{O}_{I}$ is the interaction-picture operator instead of
the original operator in Schrödinger picture.

\section{General applications}

\subsection{Time-independent Hamiltonian}

Let us first apply our general result to the simple case where the
parameters in the Hamiltonian are time-independent. That is, $\hat{H}=\hat{H}(a)$,
where $a$ represents random variable(s) instead of random process(es).
Suppose further that $\hat{H}$ is of zero-mean, which implies a particular
choice of {}``picture'': Any time-independent, deterministic part
of the Hamiltonian plus the average component of the stochastic part
can be removed by a gauge transformation, that is, by switching to
an suitably chosen interaction picture \cite{messiah}. Note that,
in the case of time-independent random variables, $\hat{U}_{1}(t,t_{0})=(t-t_{0})\hat{H}/i\hbar$
and $\hat{U}_{1}^{\dagger}(t,t_{0})=-\hat{U}_{1}(t,t_{0})$, thus
$[\hat{U}_{1},\,\hat{H}]=[\hat{U}_{1}^{\dagger},\,\hat{H}]=0$. So
we have $\hat{A}+\hat{A}^{\dagger}=0$ and thus $\hat{H}_{eff}=0$.
This result is special to the time-independent case, however. As we
will see later, the effective Hamiltonian (to 2nd order) is in general
non-zero, due to the non-commutativity of $\hat{U}_{1}$ and $\hat{H}$
in the time-dependent case. After simplification, the 2nd-order master
equation in this particular case is

\begin{equation}
\frac{\partial}{\partial t}\overline{\rho}(t)=-\frac{t}{\hbar^{2}}\{\overline{\hat{H^{2}}},\overline{\rho}(t)\}+\frac{2t}{\hbar^{2}}\overline{\hat{H}\overline{\rho}(t)\hat{H}}.\label{eq:t-indep me}
\end{equation}

A class of problems of physical interest has a Hamiltonian of the
form

\begin{equation}
\hat{H}=\hbar\underset{n}{\sum}a_{n}\hat{h}_{n}+a_{n}^{*}\hat{h}_{n}^{\dagger},\label{eq:H}
\end{equation}
 where $a_{n}$ are jointly circular complex Gaussian random variables
of zero mean.%
\footnote{According to the central limit theorem, Gaussian statistics is applicable
when the random variables are due to the addition of many uncorrelated
random sources.%
} Substituting Eq.(\ref{eq:H}) into Eq.(\ref{eq:t-indep me}), we
find%
\footnote{For those readers who might be concerned about the factor $t$ on
the right-hand side of Eq.(\ref{eq:me}): It is just a result of a
time integral, as can be seen in the more general case of time-dependent
Hamiltonians later. Note that the trace-preserving property of the
equation is guaranteed by the Lindblad form.%
}

\begin{multline}
\frac{\partial}{\partial t}\overline{\rho}(t)=t\underset{k,l}{\sum}\{\Gamma_{kl}\left(-\hat{h}_{k}\hat{h}_{l}^{\dagger}\overline{\rho}(t)-\overline{\rho}(t)\hat{h}_{k}\hat{h}_{l}^{\dagger}+2\hat{h}_{l}^{\dagger}\overline{\rho}(t)\hat{h}_{k}\right)\\
+\Gamma_{kl}^{*}\left(-\hat{h}_{k}^{\dagger}\hat{h}_{l}\overline{\rho}(t)-\overline{\rho}(t)\hat{h}_{k}^{\dagger}\hat{h}_{l}+2\hat{h}_{l}\overline{\rho}(t)\hat{h}_{k}^{\dagger}\right)\},\label{eq:me}
\end{multline}
 where $\Gamma_{kl}=\overline{a_{k}a_{l}^{*}}$ are the correlation
functions. Note that the equation is of the familiar Lindblad form,
which can be further simplified to a diagonal form through a linear
transformation of the coefficients.

\subsection{Single real Gaussian random process}

Now consider the case of a time-dependent Hamiltonian 
\begin{equation}
\hat{H}(t)=\hbar a(t)\hat{h},\label{eq:H-single Grp}
\end{equation}
 where $a(t)$ is a (real) Gaussian random process. This is representative
of a wide class of problems, for example, the Zeeman effect, where
$a(t)$ is proportional to the external magnetic field and $\hat{h}$
is the $z$-component of the total angular momentum \cite{bransden}.
The random process in this section is taken to be the most general
case, that is, we do not assume any additional property like zero-mean
or stationarity.

The ensemble-average dynamics under this Hamiltonian is exactly solvable,
so let us first work out the exact, analytic result. The unitary evolution
operator in a particular realization $\mu$ is

\begin{equation}
\hat{U}^{\mu}(t,t_{0})=\exp\left(-iv^{\mu}(t)\hat{h}\right),
\end{equation}
 where

\begin{equation}
v^{\mu}(t)\equiv\intop_{t_{0}}^{t}dt'a^{\mu}(t').\label{eq:v}
\end{equation}
 For an initial state $\rho(t_{0})=\underset{k,l}{\sum}\rho_{kl}(t_{0})|k\rangle\langle l|$,
where $\left\{ |n\rangle\right\} $ is the energy-eigenbasis with
$\hat{h}|n\rangle=E_{n}|n\rangle$, the evolution in a particular
realization is

\begin{equation}
\rho^{\mu}(t)=\hat{U}^{\mu}(t,t_{0})\rho(t_{0})\hat{U}^{\mu\dagger}(t,t_{0})=\underset{k,l}{\sum}\rho_{kl}(t_{0})\exp\left\{ -iv^{\mu}(t)(E_{k}-E_{l})\right\} |k\rangle\langle l|,
\end{equation}
 thus the ensemble-average is

\begin{equation}
\overline{\rho}(t)=\underset{k,l}{\sum}\rho_{kl}(t_{0})\overline{\exp\left\{ -iv(t)(E_{k}-E_{l})\right\} }|k\rangle\langle l|.
\end{equation}
 Invoking the special properties of Gaussian statistics,%
\footnote{Since $v(t)$ is a linear filtered Gaussian random process, it is
a Gaussian random process itself. (See page 83 of Ref.\cite{goodman}.)%
} it can be shown that

\begin{equation}
\overline{\exp\left\{ -iv(t)(E_{k}-E_{l})\right\} }=\exp\left\{ -i(E_{k}-E_{l})\overline{v(t)}-\frac{(E_{k}-E_{l})^{2}}{2}\left[\overline{v(t)^{2}}-\overline{v(t)}^{2}\right]\right\} .
\end{equation}
 Thus, the exact ensemble-average dynamics is given by the elements
of the average density matrix:

\begin{eqnarray}
\overline{\rho}_{kk}(t) & = & \rho_{kk}(t_{0}),\label{eq:exact diag}\\
\overline{\rho}_{kl}(t) & = & \rho_{kl}(t_{0})\exp\left\{ -i(E_{k}-E_{l})\overline{v(t)}-\frac{(E_{k}-E_{l})^{2}}{2}[\overline{v(t)^{2}}-\overline{v(t)}^{2}]\right\} .\label{eq:exact off-diag}
\end{eqnarray}

Now let us solve the same problem by the master equation approach.
Using the results from Eqs.(\ref{eq:effective H}-\ref{eq:general me}),
the following expression is obtained,

\begin{multline}
i\hbar\frac{\partial}{\partial t}\overline{\rho}(t)=[\hbar\overline{a(t)}\hat{h},\overline{\rho}(t)]\\
+i\hbar\left[\overline{a(t)}\intop_{t_{0}}^{t}dt'\overline{a(t')}-\intop_{t_{0}}^{t}dt'\overline{a(t)a(t')}\right]\left(\hat{h}^{2}\overline{\rho}(t)+\overline{\rho}(t)\hat{h}^{2}\right)\\
+2i\hbar\left[\intop_{t_{0}}^{t}dt'\overline{a(t)a(t')}-\overline{a(t)}\intop_{t_{0}}^{t}dt'\overline{a(t')}\right]\hat{h}\overline{\rho}(t)\hat{h},
\end{multline}
 which can be simplified to

\begin{equation}
\frac{\partial}{\partial t}\overline{\rho}(t)=-i\overline{a(t)}[\hat{h},\overline{\rho}(t)]+D(t)\left[\hat{h},[\hat{h},\overline{\rho}(t)]\right],\label{eq:me2}
\end{equation}
 where

\begin{equation}
D(t)\equiv\overline{a(t)}\intop_{t_{0}}^{t}dt'\overline{a(t')}-\intop_{t_{0}}^{t}dt'\overline{a(t)a(t')}.
\end{equation}
 To find the solution to this differential equation, we first write
it down in terms of the matrix elements in the $\hat{h}$-eigenbasis:

\begin{eqnarray}
\frac{\partial}{\partial t}\overline{\rho}_{kk}(t) & = & 0,\\
\frac{\partial}{\partial t}\overline{\rho}_{kl}(t) & = & \left[-i\overline{a(t)}(E_{k}-E_{l})+D(t)(E_{k}-E_{l})^{2}\right]\overline{\rho}_{kl}(t),\,\,(k\neq l).
\end{eqnarray}
 Now we have a set of (de-coupled) linear ordinary differential equations
(ODE's)\emph{,} which is easily solvable,

\begin{eqnarray}
\overline{\rho}_{kk}(t) & = & \rho_{kk}(t_{0}),\label{eq:diagonal}\\
\overline{\rho}_{kl}(t) & = & \rho_{kl}(t_{0})\exp\left\{ \intop_{t_{0}}^{t}dt'\left[-i\overline{a(t')}(E_{k}-E_{l})+D(t')(E_{k}-E_{l})^{2}\right]\right\} \nonumber \\
 & = & \rho_{kl}(t_{0})\exp\left\{ -i(E_{k}-E_{l})\overline{v(t)}-\frac{(E_{k}-E_{l})^{2}}{2}[\overline{v(t)^{2}}-\overline{v(t)}^{2}]\right\} .\label{eq:off-diagonal}
\end{eqnarray}
 The second equality in Eq.(\ref{eq:off-diagonal}) is obtained after
some calculation, where $v(t)$ is given by Eq.(\ref{eq:v}). Thus
we find the dynamics generated by the 2nd-order master equation \emph{coincides}
with the exact dynamics in this case.

We observe that the energy population is conserved during the evolution
while the coherence between different energy levels decays. Thus the
evolution of the average dynamics is pure decoherence, with the {}``pointer
basis'' \cite{schlosshauer} being the energy-eigenbasis. Note that,
although the Hamiltonian varies with time and across different realizations,
the energy-eigenbasis is the same throughout. In the case where some
energy level is degenerate, we readily have a {}``decoherence-free
subspace'', in which quantum information can be stably stored \cite{lidar}.
Incidentally, a derivation in the context of open systems also suggests
that energy eigenstates can emerge as {}``pointer states'' in the
so-called {}``quantum limit of decoherence'' \cite{paz}. In that
case, however, the decoherence results from small environmental perturbation
to the system, not from the fluctuation of the system Hamiltonian
itself.

We could have worked out the higher-order terms (i.e. $\mathcal{L}_{n}[\rho]$
for $n\geqslant3$) explicitly to see how accurate the 2nd-order approximation
is. However, since the solution to the 2nd-order master equation coincides
with the exact dynamics, we can readily conclude that all higher-order
terms must sum up to zero without actually carrying out further calculations.

Note that when deriving Eq.(\ref{eq:me2}) we do not assume anything
about the nature of the random process $a(t)$, not even the Gaussian
statistics. In other words, the solution to the 2nd-order master equation
is given by Eqs.(\ref{eq:diagonal}-\ref{eq:off-diagonal}) in all
cases. On the other hand, the exact dynamics Eqs.(\ref{eq:exact diag}-\ref{eq:exact off-diag})
is based on the assumption of Gaussian statistics. If $a(t)$ is \emph{not}
a Gaussian random process, then the exact dynamics will be different.%
\footnote{It may not be exactly solvable, but we know for sure that the solution
is different from that in the Gaussian case.%
} The implication is that, for $a(t)$ being non-Gaussian, the 2nd-order
master equation is not exact.

\subsection{Multiple jointly circular complex Gaussian random processes}

Let us briefly present the results for the more general Hamiltonian
$\hat{H}(t)=\hbar\underset{n}{\sum}\left(a_{n}(t)\hat{h}_{n}+a_{n}^{*}(t)\hat{h}_{n}^{\dagger}\right)$,
where $a_{n}(t)$ are jointly circular complex Gaussian random processes
of zero mean. The 2nd-order master equation in Lindblad form is found
to be

\begin{multline}
\frac{\partial}{\partial t}\overline{\rho}(t)=-\underset{k,l}{\sum}\alpha_{kl}(t)\left[[\hat{h}_{k},\hat{h}_{l}^{\dagger}],\,\overline{\rho}(t)\right]+\underset{k,l}{\sum}\beta_{kl}(t)\\
\times\left\{ -\hat{h}_{k}\hat{h}_{l}^{\dagger}\overline{\rho}(t)-\overline{\rho}(t)\hat{h}_{k}\hat{h}_{l}^{\dagger}+2\hat{h}_{l}^{\dagger}\overline{\rho}(t)\hat{h}_{k}-\hat{h}_{l}^{\dagger}\hat{h}_{k}\overline{\rho}(t)-\overline{\rho}(t)\hat{h}_{l}^{\dagger}\hat{h}_{k}+2\hat{h}_{k}\overline{\rho}(t)\hat{h}_{l}^{\dagger}\right\} ,\label{eq:me3}
\end{multline}
 where 
\begin{equation}
\alpha_{kl}(t)\equiv\frac{1}{2}\intop_{t_{0}}^{t}dt'\left(\overline{a_{k}(t)a_{l}^{*}(t')}-\overline{a_{l}^{*}(t)a_{k}(t')}\right),
\end{equation}

\begin{equation}
\beta_{kl}(t)\equiv\frac{1}{2}\intop_{t_{0}}^{t}dt'\left(\overline{a_{k}(t)a_{l}^{*}(t')}+\overline{a_{l}^{*}(t)a_{k}(t')}\right).
\end{equation}
 By comparing with Eq.(\ref{eq:me}) for the time-independent Hamiltonian
case, we notice a major difference in this case is that the effective
Hamiltonian is non-zero despite $\overline{\hat{H}(t)}=0$. This effective
unitary evolution results from the fact that the Hamiltonian operators
at different times do not commute with each other in general.

The 2nd-order master equation yields exact dynamics only for the special
case of a single real Gaussian random process. In this more general
case, Eq.(\ref{eq:me3}) does not lead to exact dynamics in general.
This can be shown by explicitly evaluating higher-order terms like
$\mathcal{L}_{4}[\rho]$ to find that they do not vanish in general.
Despite the lack of perfect agreement, the master equation is nevertheless
of great use in such cases, because the exact dynamics is generally
not obtainable and the 2nd-order master equation serves as a good
approximation when the higher-order terms (e.g. $\mathcal{L}_{4}[\rho]$)
are small compared to $\mathcal{L}_{2}[\rho]$.

\section{Physical examples}

We will illustrate the general results by applying them to a few examples.
The findings will then be used to gain physical insights, and the
validity of the master equation approach will be examined by comparing
to the exact dynamics.

\subsection{Two-level system}

First consider a two-level system (e.g. spin-1/2) subject to the Hamiltonian
$\hat{H}(t)=\hbar\omega(t)\hat{Z}$, where $\hat{Z}$ is the $z$-component
of Pauli operator and $\omega(t)$ a stationary Gaussian random process
of zero mean. This falls into the category of Hamiltonians (\ref{eq:H-single Grp}).
Using Eq.(\ref{eq:me2}), the 2nd-order master equation is obtained,

\begin{equation}
\frac{\partial}{\partial t}\overline{\rho}(t)=-\frac{1}{4}d(t)\left[\hat{Z},[\hat{Z},\overline{\rho}(t)]\right],
\end{equation}
 where $d(t)\equiv4\intop_{t_{0}}^{t}dt'\overline{\omega(t)\omega(t')}$.
Assuming an auto-correlation function of the form $\overline{\omega(t)\omega(t')}=\overline{\omega_{0}^{2}}\exp\left(-|t-t'|/T\right)$,
where $\overline{\omega_{0}^{2}}\equiv\overline{\omega(t)^{2}}$,
we have $d(t)=4\overline{\omega_{0}^{2}}T\left(1-e^{-(t-t_{0})/T}\right)$
for $t>t_{0}$.

Written in the $\hat{Z}$-eigenbasis $\left\{ |0\rangle,|1\rangle\right\} $,
the master equation becomes a set of linear ODE's:

\begin{eqnarray}
\frac{\partial}{\partial t}\overline{\rho}_{00}(t) & = & 0,\\
\frac{\partial}{\partial t}\overline{\rho}_{11}(t) & = & 0,\\
\frac{\partial}{\partial t}\overline{\rho}_{01}(t) & = & -d(t)\overline{\rho}_{01}(t),\\
\frac{\partial}{\partial t}\overline{\rho}_{10}(t) & = & -d(t)\overline{\rho}_{10}(t);
\end{eqnarray}
 the solutions to which are

\begin{eqnarray}
\overline{\rho}_{00}(t) & = & \rho_{00}(t_{0}),\\
\overline{\rho}_{11}(t) & = & \rho_{11}(t_{0}),\\
\overline{\rho}_{01}(t) & = & \rho_{01}(t_{0})\exp\left\{ -4\overline{\omega_{0}^{2}}T^{2}\left(\frac{t-t_{0}}{T}+e^{-(t-t_{0})/T}-1\right)\right\} ,\\
\overline{\rho}_{10}(t) & = & \rho_{10}(t_{0})\exp\left\{ -4\overline{\omega_{0}^{2}}T^{2}\left(\frac{t-t_{0}}{T}+e^{-(t-t_{0})/T}-1\right)\right\} .
\end{eqnarray}
 As already discussed in the general case of a single Gaussian random
process, the energy population remains constant while the coherence
decays. This can be understood from a more physical perspective. Quantum
coherence depends on the relative phase between the two components
$|0\rangle$ and $|1\rangle$. In an individual realization, the relative
phase factor is $\exp\left\{ 2i\intop_{t_{0}}^{t}dt'\omega(t')\right\} $.
Since $\omega(t)$ is a random process, the quantity $\intop_{t_{0}}^{t}dt'\omega(t')$
becomes increasingly randomized with the passage of time. When the
average is taken over an ensemble, these randomly distributed relative
phase factors cancel out, thus suppressing the coherence. This suggests
that quantum interference is difficult to observe because random fluctuation
is ubiquitous.

As has been shown in the more general case Eqs.(\ref{eq:diagonal}-\ref{eq:off-diagonal}),
the 2nd-order master equation gives exact dynamics. When we work out
the exact dynamics directly, the result is indeed found to be consistent,
though such a direct calculation is more demanding. Clearly, calculational
convenience is one advantage of the master equation approach.

\subsection{A pair of two-level systems in magnetic field}

Next consider an example of two atoms in magnetic field, each atom
being a two-level system. The interaction of the spin with the B-field
is given by $\hat{H}(t)=\hbar\Omega(t)\left(\hat{Z}^{I}\otimes\hat{I}^{II}+\hat{I}^{I}\otimes\hat{Z}^{II}\right)\equiv\hbar\Omega(t)\hat{Z}_{total}$,
where $\hat{Z}^{j}$ is the usual Pauli $z$-operator of the $j$-th
atom. Suppose that the frequency $\Omega(t)$, which is proportional
to the B-field strength, is a stationary Gaussian random process of
zero mean and that its auto-correlation is of a Markovian type $\overline{\Omega(t)\Omega(t')}=\frac{1}{8}\gamma\delta(t-t')$,
where the constant $\gamma$ has dimension of inverse-time. Applying
Eq.(\ref{eq:me2}), it can be shown that the 2nd-order master equation
is

\begin{equation}
\frac{\partial}{\partial t}\overline{\rho}(t)=-\frac{1}{16}\gamma\left[\hat{Z}_{total},[\hat{Z}_{total},\overline{\rho}(t)]\right].\label{eq:me-2 atoms}
\end{equation}

Suppose the system starts in an entangled state between two atoms
$|\Psi(t_{0})\rangle=\frac{1}{\sqrt{2}}\left(|01\rangle+|10\rangle\right)$.
Since it is an eigenstate of $\hat{Z}_{total}$, the right-hand side
of Eq.(\ref{eq:me-2 atoms}) is identically zero, thus the system
does not evolve (except possibly to an unobservable global phase).
So the two atoms remain entangled over time. Indeed, notice that $|01\rangle$
and $|10\rangle$ are degenerate eigenstates with the same energy.
Thus any arbitrary superposition state of $|01\rangle$ and $|10\rangle$
will remain unchanged over time; in particular, the coherence between
them does not decay. Thus, any state in this degenerate subspace is
immune to decoherence, making it a good place to store quantum information
\cite{lidar}.

Let us see what happens if the 2-atom system starts in a different
entangled state like $|\Psi(t_{0})\rangle=\frac{1}{\sqrt{2}}\left(|00\rangle+|11\rangle\right)$.
Writing the master equation in the energy eigenbasis $\left\{ |00\rangle,|01\rangle,|10\rangle,|11\rangle\right\} $,
we obtain a set of decoupled linear ODE's as usual. The solutions
are found to be

\begin{eqnarray}
\overline{\rho}_{00,00}(t) & = & \overline{\rho}_{11,11}(t)=\frac{1}{2},\\
\overline{\rho}_{00,11}(t) & = & \overline{\rho}_{11,00}(t)=\frac{1}{2}\exp\left\{ -\gamma(t-t_{0})\right\} ,
\end{eqnarray}
 while the rest of the matrix elements are identically zero. Note
that decoherence occurs here, as is expected, since the initial state
does not lie in the decoherence-free subspace. Furthermore, as $t\rightarrow\infty$,
the coherence is suppressed to zero and $\overline{\rho}(t)\rightarrow\frac{1}{2}|00\rangle\langle00|+\frac{1}{2}|11\rangle\langle11|=\frac{1}{2}|0\rangle\langle0|\otimes|0\rangle\langle0|+\frac{1}{2}|1\rangle\langle1|\otimes|1\rangle\langle1|$.
Interestingly, the two atoms become \emph{disentangled}, as there
is no quantum correlation between them. In contrast to the general
belief that entanglement leads to decoherence, as is widely studied
for open quantum systems, here we find that decoherence can result
in\emph{ }disentanglement\emph{ }in the case of a closed system.

\subsection{Heating of a trapped ion}

Consider an ion with mass $M$ and charge $e$ in a harmonic binding
potential with characteristic frequency $\omega_{0}$. The ion is
driven by a classical electric field $E(t)$, which is a stationary
Gaussian random process of zero mean. It is more convenient to work
in the interaction picture, in which the easily solvable, deterministic
evolution induced by the harmonic potential is treated separately.
The interaction-picture Hamiltonian%
\footnote{Throughout this section we work in the interaction picture. The subscripts
to denote interaction-picture operators are dropped for notational
simplicity.%
} is given by $\hat{H}(t)=i\hbar\left[u(t)\hat{a}^{\dagger}-u^{*}(t)\hat{a}\right]$,
where $u(t)=ieE(t)e^{i\omega_{0}t}/\sqrt{2M\hbar\omega_{0}}$ and
$\hat{a}$ $(\hat{a}^{\dagger})$ being the zero-time annihilation
(creation) operator for the harmonic oscillator. The evolution is
exactly solvable and the analytic results are given in \cite{james}.

Let us derive the 2nd-order master equation for this case. Note that
it does not fall into the category of a single real Gaussian random
process as in Eq.(\ref{eq:H-single Grp}). Since $\overline{\hat{H}}=0$,
applying Eq.(\ref{eq:effective H}), the effective Hamiltonian is
found to be 
\begin{eqnarray}
\hat{H}_{eff} & = & \frac{1}{2}\intop_{t_{0}}^{t}dt'\left(\overline{u(t)u^{*}(t')}-\overline{u^{*}(t)u(t')}\right)[\hat{a},\hat{a}^{\dagger}]\nonumber \\
 & = & -\frac{e^{2}}{2M\omega_{0}}\intop_{t_{0}}^{t}dt'\overline{E(t)E(t')}\sin\left[\omega_{0}(t-t')\right]\hat{I}.
\end{eqnarray}
 In this case, the 2nd-order contribution to the effective Hamiltonian
is non-zero, a consequence of the non-commutativity of $\hat{H}(t)$
and $\hat{U}_{1}(t)$. However, since $\hat{H}_{eff}$ is proportional
to $\hat{I}$, the unitary part of the equation of motion results
only in an unobservable global phase in this case. For more general
cases, $\hat{H}_{eff}$ can be different from the identity $\hat{I}$
and can well lead to non-trivial dynamics. Evaluating the remaining
terms in Eq.(\ref{eq:L(2)}) for this example, we find the following
master equation:

\begin{multline}
\frac{\partial}{\partial t}\overline{\rho}(t)=-\mathcal{C}(t)\left(\hat{a}^{\dagger}\hat{a}\overline{\rho}(t)+\overline{\rho}(t)\hat{a}^{\dagger}\hat{a}-2\hat{a}\overline{\rho}(t)\hat{a}^{\dagger}\right)\\
-\mathcal{C}(t)\left(\hat{a}\hat{a}^{\dagger}\overline{\rho}(t)+\overline{\rho}(t)\hat{a}\hat{a}^{\dagger}-2\hat{a}^{\dagger}\overline{\rho}(t)\hat{a}\right)\\
-e^{2i\omega_{0}t}\left[\mathcal{C}(t)-i\mathcal{S}(t)\right]\left((\hat{a}^{\dagger})^{2}\overline{\rho}(t)+\overline{\rho}(t)(\hat{a}^{\dagger})^{2}-2\hat{a}^{\dagger}\overline{\rho}(t)\hat{a}^{\dagger}\right)\\
-e^{-2i\omega_{0}t}\left[\mathcal{C}(t)+i\mathcal{S}(t)\right]\left(\hat{a}^{2}\overline{\rho}(t)+\overline{\rho}(t)\hat{a}^{2}-2\hat{a}\overline{\rho}(t)\hat{a}\right),
\end{multline}
 where $\mathcal{C}(t)$ ($\mathcal{S}(t)$) are proportional to the
incomplete cosine (sine) transform of the field correlation function,
viz 
\begin{eqnarray}
\mathcal{C}(t) & \equiv & \frac{e^{2}}{2M\hbar\omega_{0}}\intop_{t_{0}}^{t}dt'\overline{E(t)E(t')}\cos\left[\omega_{0}(t-t')\right],\\
\mathcal{S}(t) & \equiv & \frac{e^{2}}{2M\hbar\omega_{0}}\intop_{t_{0}}^{t}dt'\overline{E(t)E(t')}\sin\left[\omega_{0}(t-t')\right].
\end{eqnarray}
 Assuming $\overline{E(t)E(t')}=\overline{E(0)^{2}}\exp\left(-|t-t'|/T\right)$
and setting $t_{0}=0$ for convenience, we find $\mathcal{C}(t)=\left(1/2\tau_{1}\right)\left\{ e^{-t/T}\left[\omega_{0}T\sin(\omega_{0}t)-\cos(\omega_{0}t)\right]+1\right\} $
and $\mathcal{S}(t)=-\left(1/2\tau_{1}\right)\left\{ e^{-t/T}\left[\sin(\omega_{0}t)+\omega_{0}T\cos(\omega_{0}t)\right]-\omega_{0}T\right\} $,
where $\tau_{1}$ is the heating time defined as $1/\tau_{1}=\left(e^{2}\overline{E(0)^{2}}/M\hbar\omega_{0}\right)\left(T/(1+\omega_{0}^{2}T^{2})\right)$.

Unlike the case of Eq.(\ref{eq:H-single Grp}), the 2nd-order master
equation does not generate exact dynamics in this case. To get an
approximation of the heating from the ground state (i.e. $\rho_{00}(0)=1$)
for a short period of time, let us write the master equation in the
energy eigenbasis of the harmonic oscillator,

\begin{multline}
\frac{\partial}{\partial t}\overline{\rho}_{00}(t)=-2\mathcal{C}(t)\overline{\rho}_{00}(t)+2\mathcal{C}(t)\overline{\rho}_{11}(t)\\
-\sqrt{2}e^{2i\omega_{0}t}\left[\mathcal{C}(t)-i\mathcal{S}(t)\right]\overline{\rho}_{02}(t)-\sqrt{2}e^{-2i\omega_{0}t}\left[\mathcal{C}(t)+i\mathcal{S}(t)\right]\overline{\rho}_{20}(t).
\end{multline}
 Since $\overline{\rho}_{11}(t)$, $\overline{\rho}_{02}(t)$ and
$\overline{\rho}_{20}(t)$ are all negligibly small for $t\ll T,\,1/\omega_{0}$,
we have $\frac{\partial}{\partial t}\overline{\rho}_{00}(t)\cong-2\mathcal{C}(t)\overline{\rho}_{00}(t)$
to the lowest order. In the same manner, since the depopulation $1-\overline{\rho}_{00}(t)$
is perturbatively small, we have $\overline{\rho}_{00}(t)\cong1$
to the lowest order on the right-hand side. Thus an approximate differential
equation is obtained as $\frac{\partial}{\partial t}\overline{\rho}_{00}(t)\cong-2\mathcal{C}(t)$.
Solving this ODE, we find, to lowest order,%
\footnote{The same result can be obtained by working out the evolution of $\overline{\rho}_{11}(t)$
for $t\ll T,1/\omega_{0}$ using approximation to the same order.%
}

\begin{equation}
1-\overline{\rho}_{00}(t)\cong2\intop_{0}^{t}dt'\mathcal{C}(t')\cong\frac{e^{2}\overline{E(0)^{2}}}{2M\hbar\omega_{0}}t^{2},
\end{equation}
 which holds for short times and agrees with the analytic result in
\cite{james}.

To investigate the evolution of the system for longer times, we write
down the master equation in the same basis and solve it numerically.
Since the Hilbert space is of infinite dimensions, it is not possible
to write down the complete set of ODE's for the matrix elements. Instead,
we truncate it to a set of $5\times5$ coupled ODE's that includes
only the matrix elements of the five lowest energy-eigenstates and
their coherence.%
\footnote{Since the system is of continuous nature, we could have solved the
master equation in the Wigner representation. However, that approach
is not analytically solvable either and is not computationally economical.
Furthermore, even in that case, we still have to accept the imperfection
of truncation since the numerics can only be done on a finite region
of the {}``phase space''.%
} The numerical solutions of $F(t)\equiv\overline{\rho}_{00}(t)$ (i.e.
fidelity of the ground state) are shown in Figure 1 for different
sets of parameters. It can be seen that, as $\omega_{0}\tau_{1}$
(i.e. the dimensionless heating time) increases with $\omega_{0}T$
(i.e. the dimensionless coherence time of $\overline{E(t)E(t')}$)
fixed, the numerical result gives better approximation to the exact
dynamics. Also note that, for larger values of $\omega_{0}T$, the
ground state population shows temporary revival against its general
trend of decrease.

\begin{figure}
\includegraphics[width=6.5cm]{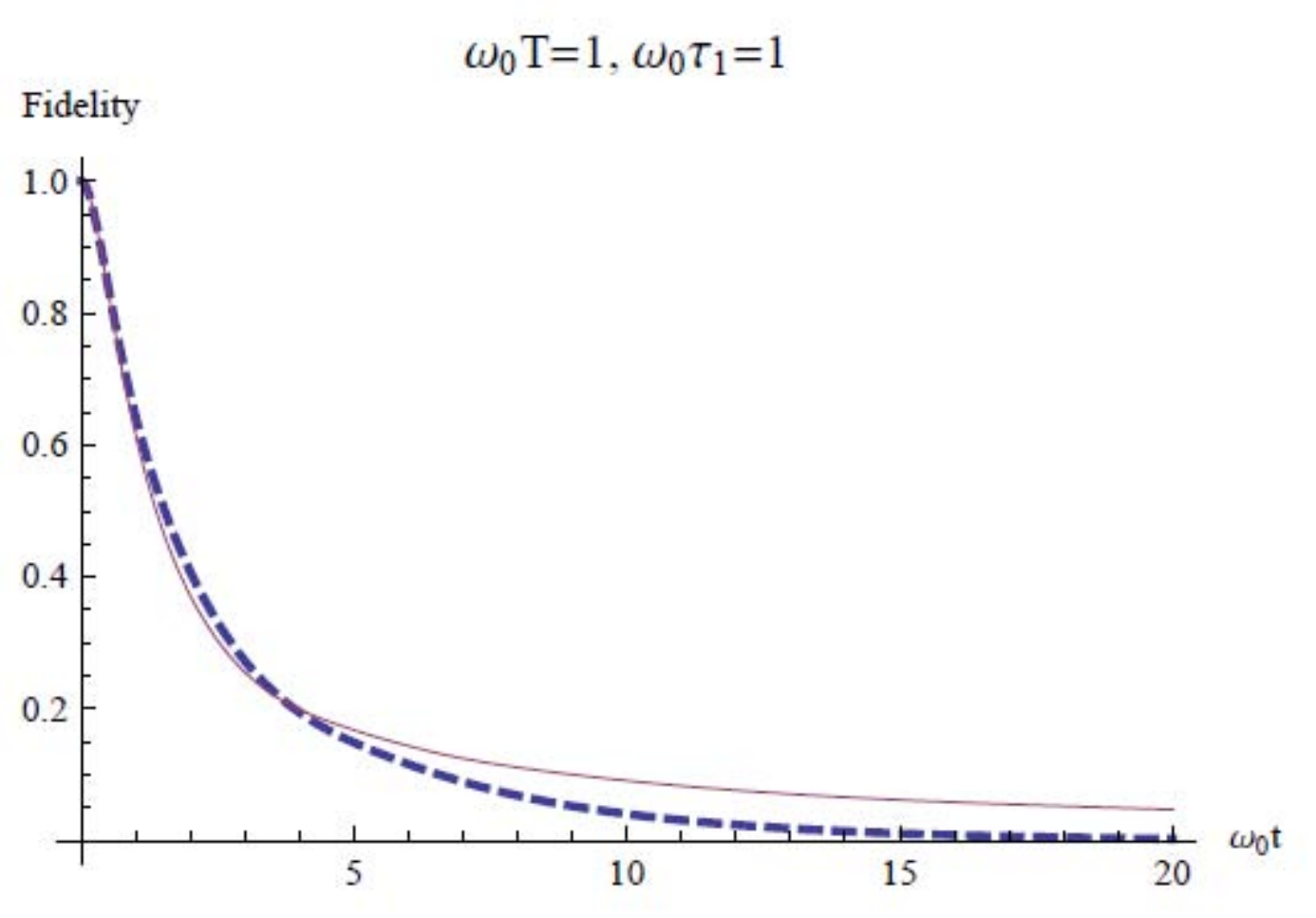}\includegraphics[width=6.5cm]{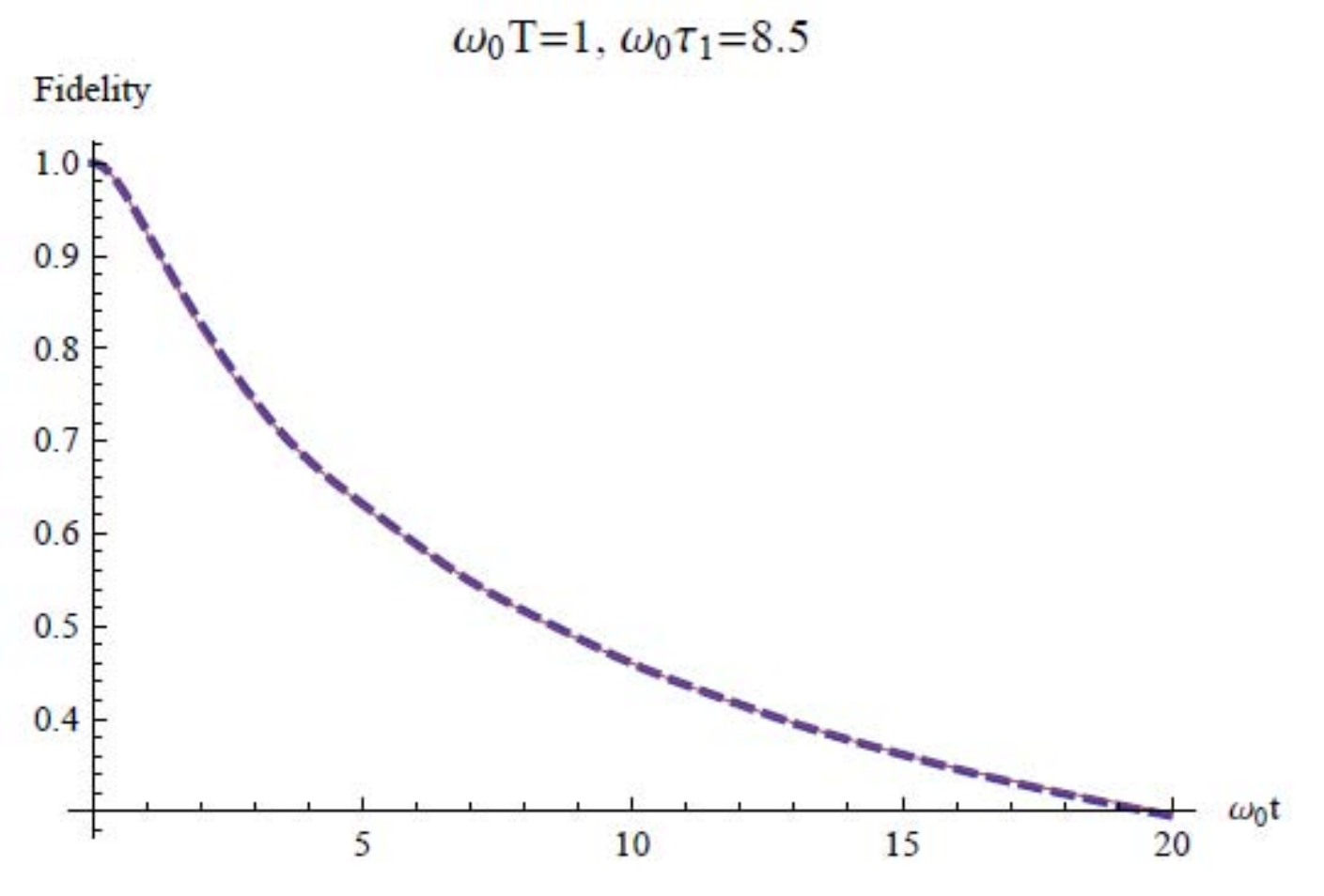}

\includegraphics[width=6.5cm]{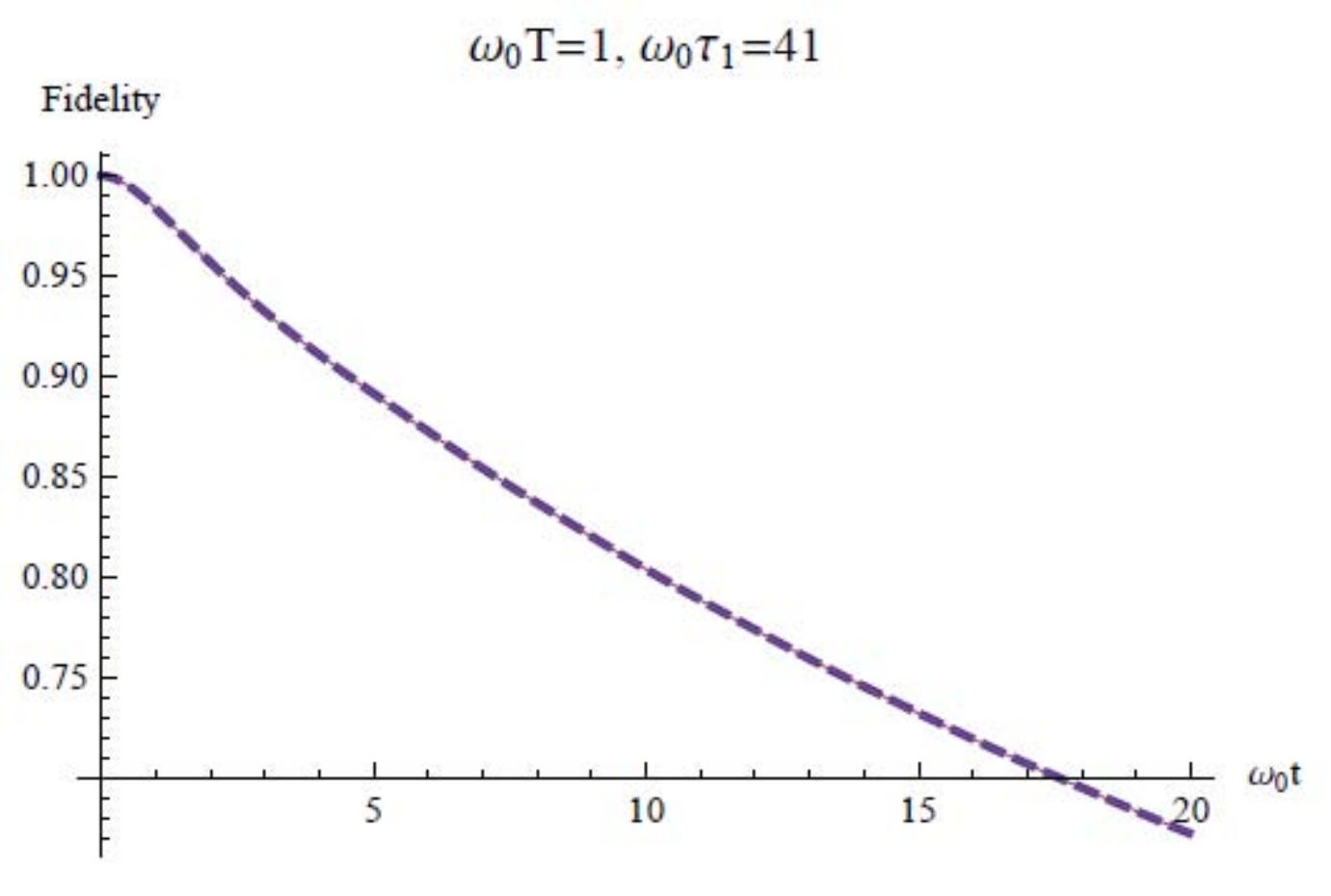}\includegraphics[width=6.5cm]{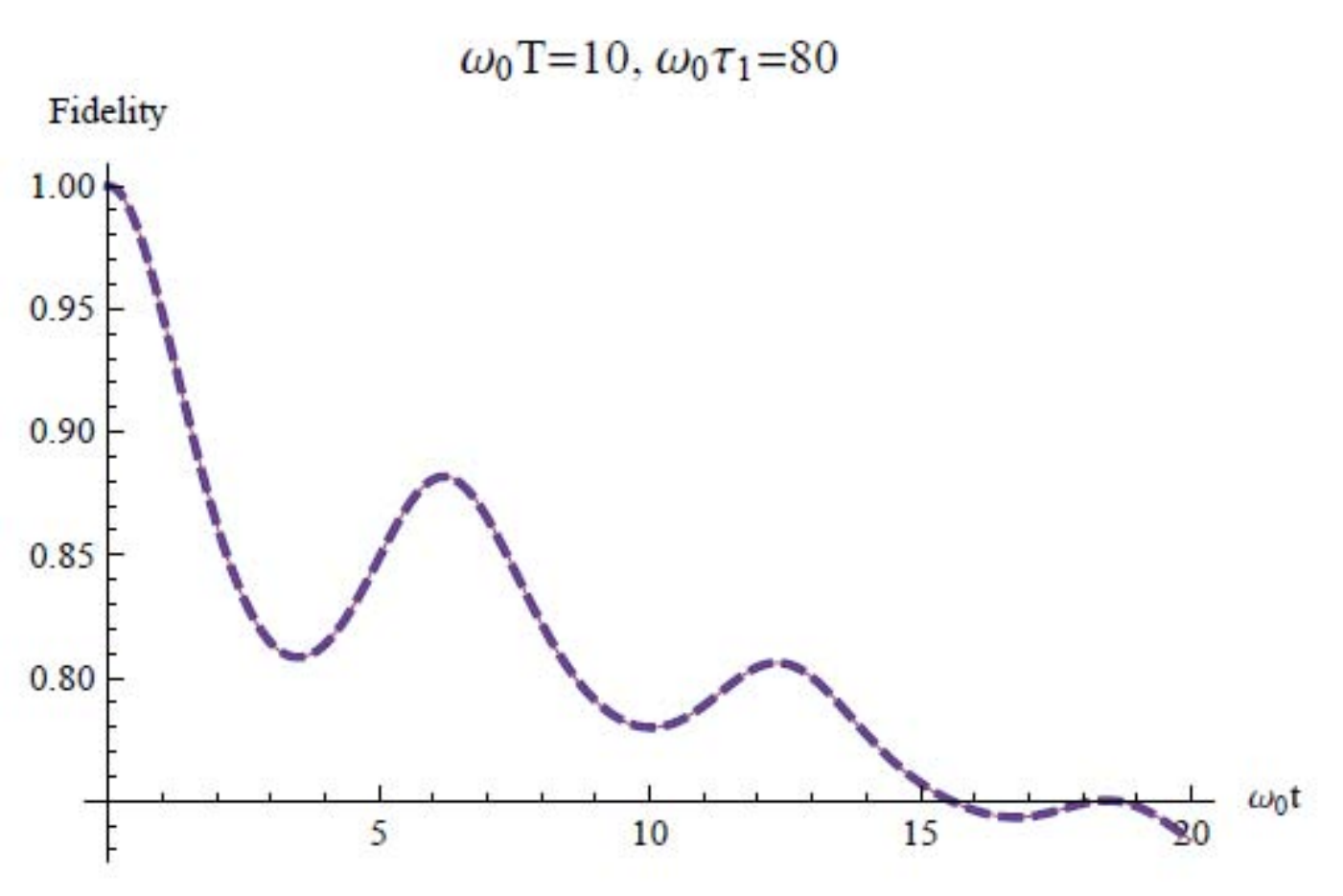}

\caption{\textsf{\footnotesize The fidelity of the ground state as a function
of dimensionless time }\textsf{\textbf{\footnotesize $\omega_{0}t$}}\textsf{\footnotesize .
Dash lines represent our numerical results, while solid lines are
exact dynamics from \cite{james}.}}
\end{figure}

Despite the artificial defect caused by the truncation of the set
of ODE's, it is of more interest to know the validity of the 2nd-order
master equation itself in approximating the exact dynamics. This is
done by comparing the size of higher-order terms to that of the 2nd-order
term. Using the Gaussian moment theorem\cite{goodman}, it is easy
to show that $\mathcal{L}_{n}[\rho]=0$ for all odd numbers $n$,
so we are interested in the ratios between the even-number-order terms.
Assuming $\omega_{0}T$ is fixed, it can be shown that $\mathcal{L}_{4}[\rho]\propto1/\tau_{1}^{2}\omega_{0}$
as opposed to $\mathcal{L}_{2}[\rho]\propto1/\tau_{1}$, so $\mathcal{L}_{4}[\rho]/\mathcal{L}_{2}[\rho]\propto1/\omega_{0}\tau_{1}$.
The same ratio holds for $\mathcal{L}_{6}[\rho]/\mathcal{L}_{4}[\rho]$,
etc. Therefore, as long as $1/\omega_{0}\tau_{1}$ is small, the higher-order
terms become progressively small, lending legitimacy to the 2nd-order
approximation. This is also consistent with the previous observation
from the numerical results. Physically, this can be better understood
by switching to the Schrödinger picture: The external field $H_{field}\propto1/\tau_{1}$
is treated as a perturbation to the self-Hamiltonian of the system
$H_{self}\propto\omega_{0}$. Naturally, as the relative size of the
perturbing Hamiltonian $H_{self}/H_{field}\propto1/\omega_{0}\tau_{1}$
becomes smaller, a perturbative method such as the 2nd-order master
equation gives better approximation to the exact dynamics.

\section{Conclusion}

In this paper we have presented the derivation of a master equation
for closed systems driven by stochastic Hamiltonians from an ensemble-average
perspective. The principal result is given in Eqs.(\ref{eq:effective H}-\ref{eq:general me}).
The validity of this approach is examined and 2nd-order master equation
is found to yield either exact dynamics or good approximations to
exact dynamics.

Applying the formalism to various physical examples, we find the ensemble-average
dynamics usually contains decoherence terms in addition to the unitary
evolution. Decoherence plays an important role in the foundational
problems of quantum mechanics, as it gives insights in two aspects
of the measurement problem, namely the absence of observable superposition
and the problem of preferred basis \cite{schlosshauer}. Extensive
research has been done on how environmental entanglement causes decoherence
in open systems. However, as our findings suggest, decoherence could
also be attributed to the random fluctuations of physical quantities
in closed systems. If this is true, then the tension between the classicality
of our experience and the quantumness of the underlying laws of physics
could be reconciled in some degree by the ubiquitous random fluctuations.
Further investigation is needed to find out (a) to what extent decoherence
is actually caused by random fluctuations and (b) whether/how we can
distinguish it from the usual entanglement-induced decoherence through
physical observation.

\section*{Acknowledgements}

The authors would like to thank O. Gamel for valuable discussions
and C.-H. Chang for comments on the manuscript. This work is supported
by Natural Sciences and Engineering Research Council of Canada (NSERC)
through CREATE and by University of Toronto through UTEA-NSE.

\end{document}